# Making Sense of Knowledge Intensive Processes: an Oil & Gas Industry Scenario

**Juliana Jansen Ferreira**

**Vinícius Segura**

**Ana Fucs**

Visual Analytics & Comprehension

IBM Research, Brazil

Avenida Pasteur, 138 - Urca Rio de Janeiro - RJ, 22290-240
{jjansen, vboas, anafucs} @br.ibm.com

**Rogério de Paula**

Visual Analytics & Comprehension

IBM Research, Brazil

Rua Tutóia, 1157, Paraíso São Paulo - SP, 04007-005
ropaula@br.ibm.com

## Abstract
Sensemaking is a constant and ongoing process by which people associate meaning to experiences. It can be an individual process, known as abduction, or a group process by which people give meaning to collective experiences. The sensemaking of a group is influenced by the abduction process of each person about the experience. Every collaborative process needs some level of sensemaking to show results. For a knowledge–intensive process, sensemaking is central and related to most of its tasks. We present findings from a fieldwork executed in knowledge-intensive process from the Oil & Gas industry. Our findings indicated that different types of knowledge can be combined to compose the result of a sensemaking process (*e.g.* decision, the need for more discussion, etc.). This paper presents an initial set of knowledge types that can be combined to compose the result of the sensemaking of a collaborative decision-making process. We also discuss ideas for using systems powered by Artificial Intelligence to support sensemaking processes.

## ACM Classification Keywords
Social and professional topics → Socio-technical systems; Human-centered computing → Empirical studies in collaborative and social computing

## Introduction
According to Weick and Meader ([7], p. 232) sensemaking is the process of constructing "moderately consensual definitions that cohere long enough for people to be able to infer some idea of what they have, what they want, why they cannot get it, and why it may not be worth getting in the first place." In professional practices, it is commonly related to decision-making processes [6, 9], which directly affect business success. Several practices have challenges regarding sensemaking. Software development, for example, is a very social-technical process which sensemaking goes from the early stages (*e.g.* requirement elicitation) through software deploy and homologation. The development team among themselves [8] and with users are constantly looking to reach a common ground about innumerous impasses.

For knowledge-intensive processes, sensemaking appears even during the definition of the process itself. For instances of these processes, the sensemaking is central. All people involved in that process need to be aligned and have a common understanding about the sensemaking process results. Our sensemaking scenario was identified during a fieldwork in an Oil & Gas company. We collected the data from 9 interviews with interpreters with different formations, backgrounds, and industry experiences. Our findings unveiled different types of knowledge that comprise their sensemaking process. Those types of knowledge may help to identify the people who should be involved in the sensemaking process to achieve that company's goal. Our findings also suggest ways of using Artificial Intelligence (AI) powered systems in support of sensemaking processes.

In this position paper, we present the related work for sensemaking in knowledge-intensive contexts. Then, we describe our fieldwork. After that, we present our findings from the analysis of fieldwork data. And, in the last section, we discuss the findings, our ongoing related research, and next steps.

## Related work
The sensemaking process has strong social characteristics when, for instance, people face obstacles and impasses [2]. In attempting to overcome them, people may try to answer a set of tacit questions, such as, what is stopping me, what I can do about it, and where I can find assistance in choosing and taking an action [1]. In the context of decision-making, which may present consequences related to how people surpass those obstacles and impasses, that sensemaking may be more sensitive (*e.g.* decisions regarding people security, large amount of money, etc.).

Schön [6] says that sensemaking processes in professional practices are motivated by scenarios of complexity, instability, and uncertainty, combined with indeterminacies and value conflicts. Uncertainty and the need to making decision are characteristics of such scenarios [9]. The sensemaking of a group is influenced by each person's abduction process about the scenario and the decision to be made. Abduction is the process of forming explanatory hypotheses [4]. It is important for the study of meanings people assign to any experience because it describes the logic of human sensemaking, from practical mundane situations to more elaborate argumentation [3].

Sensemaking is intrinsic to software design and development processes. People need to make sense of a problem context, then design and build a solution for that problem, which can be, for example, a piece of software. The design rationale tells the story of that process. From problem to solution, all involved people participate in sensemaking so as to attend a client' requirement. Designers use different support tools (*e.g.* whiteboard, flipcharts, wireframes, storyboards, etc.) to register a common understanding [8] and communicate the results of each step of their design sensemaking process. Software developers externalize their sensemaking relative to the users' problems in different artifacts throughout the development process. They create documents, conceptual models, lines of code, etc. to communicate their understanding to other developers and also to end-users [3].

Software development scenario presents somehow a structured process, but when the scenario is associated with a knowledge-intensive process, the sensemaking plays the role of the process itself. A knowledge-intensive process is characterized by activities that cannot be planned in advance, may change on the fly and are driven by the contextual scenario in which the process is embedded. Who should be involved and who should be the right person to execute a particular step are dictated by the scenario. Plus, the set of users involved may not be formally defined and may be identified as the process scenario unfolds [5].

**Our sensemaking fieldwork scenario**

Our fieldwork to discuss sensemaking scenario is a knowledge intensive processes related to the seismic interpretation[1] in an Oil & Gas company. Seismic interpretation is a central process in the Exploration & Production industry and its main goal is support other decision-making processes by reducing uncertainty. To achieve that goal, different people engage in multiple informal interactions and collaboration sessions, embedding biases, decisions, and reputation. Seismic interpretation is the process of inferring the geology of a region at some depth from the processed seismic survey. A seismic survey is a data set of soundwaves refracted and reflected through Earth's crust measured and recorded respect from a particular area of the Earth's surface, to evaluate the subsurface[2]. Figure 1 and Figure 2 show examples of seismic data lines, which is a portion of a seismic survey.

---

[1] http://wiki.aapg.org/Seismic_interpretation

[2] http://www.glossary.oilfield.slb.com/Terms/s/seismic_survey.aspx

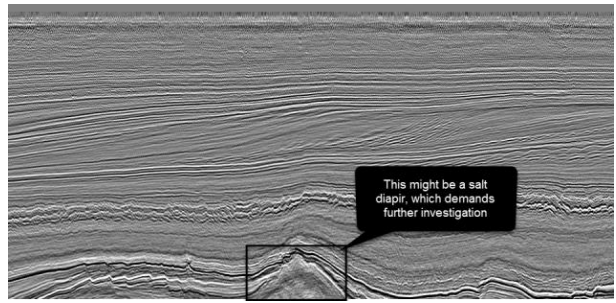

Figure 1. Seismic image example (Netherland - Central Graben – inline 474)[3]

To perform a seismic interpretation, the interpreter (which can be a geologist or a geophysicist) gets the seismic image and analyzes it based on a set of other knowledges, from papers, reference books, knowledge from previous projects, etc. In our sensemaking scenario, we have a fictional interpreter called Paul. He analyzes the seismic image and identifies a particular geological formation called salt diapir (Figure 1). He knows that this kind of formation is frequently related to regions that present potential for reservoir of hydrocarbonate, which means presence of oil or gas that can be explored and provide revenue.

After raising the hypothesis of a salt diapir existence, Paul plays with seismic data using different filters and image representation (Figure 2) in different tools (*e.g.* Petrel[4], Paradigm[5], etc.) to check if there are more visual evidences of that formation in the seismic data.

He collects more evidence to show to his peers in a review meeting.

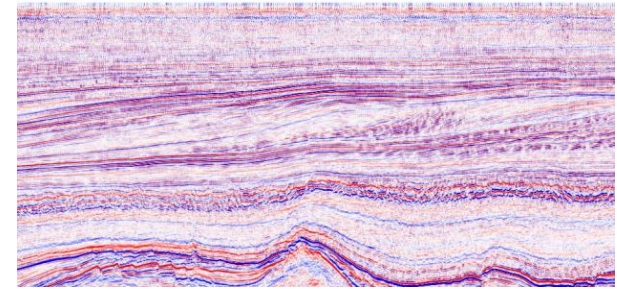

Figure 2. Same seismic line as Figure 1 with different visualization characteristic[3]

To support his salt diapir hypothesis, Paul looks for papers relating the Central Graben basin in the North Sea to the salt diapir geological formation. He finds papers, such as, "*Geometry and late-stage structural evolution of Central Graben salt diapirs, North Sea*"[6], which support his hypothesis of salt presence, with top salt indication in Figure 3.

Paul combines what he discovered in his interpretation with the knowledge he acquired from the experiences of previous projects, documents, and his personal notes. He also talks to Gary, another interpreter, who worked in another project on the Central Graben basin a few years ago, to collect more information about that basin. Once Paul is confident about his salt diapir hypothesis, he puts together a report to discuss with his peers. He prepares a presentation with his assessment about investing money on Central Graben basin area, showing all his steps to get to the hypothesis and the related evidence of return-of-investment.

---

[3] Images taken from the Open Seismic Repository, a free and public data set, of the Netherlands Offshore F3 Block - https://opendtect.org/osr /

[4] https://www.software.slb.com/products/petrel

[5] http://www.pdgm.com/solutions/seismic-processing-and-imaging/

[6] I Davison, I Alsop, P Birch, et al. 2000. Geometry and late-stage structural evolution of Central Graben salt diapirs, North Sea. Marine and Petroleum Geology 17, 4: 499–522.

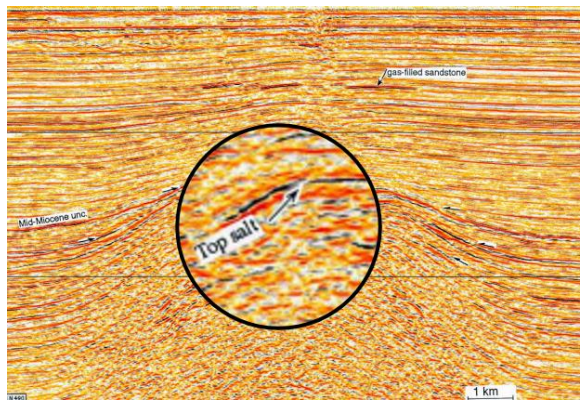

Figure 3. Seismic line interpretation presented in paper about salt diapirs in Netherland - Central Graben[6]

In the review meeting, there are a group of interpreters with different backgrounds, experiences, and formations (geologist or geophysicists). Paul presents his findings, illustrating and sustaining his hypothesis with a set of evidences. All interpreters participating in the review meeting need to get into a common ground about Paul's findings in Central Graben basin. They must decide if the company should invest in that area or not. That decision is associated with the spending of millions of dollars. They want to invest in areas that will eventually have large revenue, generating company's business value.

We anonymized the fieldwork data to protect our participant's identities and any confidential information or data. This did not impact our findings or discussions.

### Findings related to sensemaking

In the fieldwork, we identified two moments where the sensemaking needs to be externalized in some way: 1) when Paul puts together a report to be reviewed by his peers (individual sensemaking), and 2) when all interpreters involved in the review need to decide about the investment in the investigated area (collective sensemaking). This scenario in the Oil & Gas industry reveals a significant and distinct system of collaboration. Given the intrinsic risk and uncertainty properties of seismic interpretation, we observed that combining efforts in a teamwork is a way to construct a better supported model and, in a simplified vision, share responsibility over decisions in an Oil & Gas project.

Throughout the interpretation process, the sensemaking is built with different types of knowledge. The analyzed data from fieldwork allowed us to define three types of knowledge necessary in the process. These knowledges combined construct the substantial asset of information used in the seismic interpretation process. The types of knowledge are:

**Domain knowledge** – It concerns the information someone knows about the activity he/she executes in the project. For instance, an interpreter has a domain knowledge of seismic image interpretation.

**Context knowledge** – Some people might have a vast experience in a specific context, such as a country, a region, a field or a basin. This knowledge enables some inference of geological properties in a project.

**Historical knowledge** – In the Oil & Gas industry it is not unusual to observe experts that resort to previous similar projects and apply prior learnings and knowledge while creating a new model. The history of someone's career in this industry matters in a very significant way.

People with different profiles regarding types and levels of knowledge are responsible of building data resources for a project. This data stands for a result of what we call "collaborative knowledge construction" and intends to minimize uncertainties related to the project. Participants in our fieldwork highlighted this "collaborative knowledge construction":

*It always happens as a symbiosis. I'm not here working statically. I work with a geologist to have the information, to know what's his opinion, what he thinks about my project. 90% of the time the discussion ends with these people: myself, Tom, Mary, and Mike. The geologist is also always present. (Participant #1)*

*We work very close to a lot of them (interpreters). They are geophysicist... petrophysicist too...very close of all! We rely on them all. We integrate. (Participant #2)*

The quotes above show the cooperative aspect of the seismic interpretation and model creation process. Discussions, consulting, and feedback are some of the systematically identified routines. We identified some interesting characteristics in that sensemaking process of collaborative knowledge construction and categorized them as showed in Figure 4.

As we described, individualities are a significant part of a team's construction. As a teammate starts working in a project, his/her inputs contribute to the workflow of data and expands the project's resources (sensemaking) according to its knowledge type: domain, context, or historical. For instance, a less experienced interpreter (Figure 4, A) may consult a colleague with context knowledge (Figure 4, B) to gather more significant data about the basin in which his project is situated. Other colleagues with historical knowledge might collaborate (Figure 4, C, D) contributing to a much more structured information network that will grow dynamically, resulting in a lower risk decision making situation.

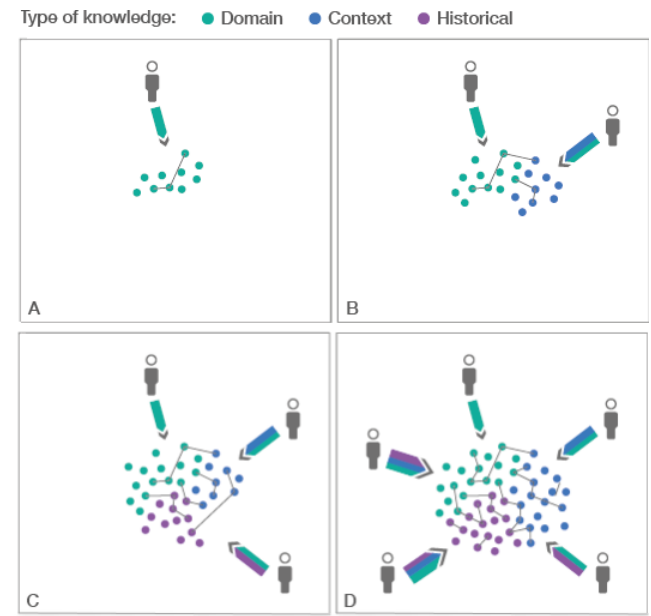

Figure 4. Collaborative Knowledge Construction

We confirm that one person might be required or demanded to participate in a team, considering all aspects of his/her professional profile. A specific experience or past participation in the same or analogue region is a strong reason for a contribution. Findings also defined profile needs. Specific experiences establish key players in the process, as we might observe in the excerpts below:

*We ask for a participant according to what was found… types of gases, fluids etc. (Participant 3)*

*Although they (other team) know how to calculate volume data conversions, they might sometimes not have sensitivity to realize if values are correct (Participant #1)*

*I did take part in several studies so I already have many case studies in my head (Participant #4)*

The whole sensemaking process in our scenario is shrouded in uncertainty, so all stages need to be composed of pieces of evidence, which must be traceable at any time to support someone else's decision or next step:

*When I receive the information, if it is not properly audited or if we can't track it, I take a risk using these parameters. Tracking the work is very important for us. This will be successively accumulated (Participant #2)*

*Let's suppose a colleague has used a cutoff he considered the best choice for that project. Then I start using a cutoff that doesn't correspond at all with the first one. For instance, if the rock does not have a minimum porosity of 5%, it is so closed up that it doesn't interests us. A petrophysicist said the cutoff was 5%. So, I realize with additional data that we need a new cutoff of 10% because 5% is not enough. There are adjustments to make. If this is not reported and tracked, we lose information and it has a direct impact on the final volume we calculate. (Participant #2)*

During the sensemaking process, the execution and its results should consistently be reviewed and adjusted according to the accumulated knowledge. Quality check is an established practice that clearly reduces risk along the project.

*We always try to set some criteria within a project. If it's wrong it needs to stand consistently wrong so it's much easier to get it corrected in the future. (…) When analyzing uncertainty, we evaluate the volumes and we have 4 to 5 controlling parameters. Our job is always to integrate more data to reduce uncertainty. (Participant #2)*

**Findings related to sensemaking**

The size of the Oil & Gas company (not a big company) influences the close collaborative interaction among people. It helps the sensemaking process by knowing who to look for if a certain knowledge is needed in a project. Some have experience in different phases in the process, so those have an overall idea of the sensemaking necessary for making a decision in a project.

The most interesting quotes were from people in the middle of the process, those who rely heavily on other people assessment (individual sensemaking) to make their own evaluation and send it to someone else. For those people, tracing the data is very important. The sensemaking process register is a huge and crucial challenge.

Different profiles act on different parts in the process. Details about people's profile (geologist, petrophysicist, etc.) interfere in the sensemaking process. The identification of key people in a sensemaking process might increase the success of the process result.

### Remarks and future work
Our fieldwork was executed in a small Oil & Gas company. In that context, people somehow know who to look for in specific situations or decisions. Also, due to the small company context, people have an idea in which project colleagues are involved in that moment. However, the current context of a company does not give any indication about a person previous experience in other companies (**historical knowledge**), with other geological contexts (**context knowledge**), or other phases in the Oil & Gas decision-making process (**domain knowledge**). That knowledge could enrich the sensemaking of future projects by involving people with the necessary knowledge to make better decisions.

Our findings indicated the need of tracking resources to support the sensemaking process for decision-making at Oil & Gas industry. People rely on others sensemaking results to build their own. To trust the input for their individual sensemaking, people need to verify the sensemaking traces of that input.

We are currently studying how Artificial Intelligence (AI) Powered Systems could impact the sensemaking process in a knowledge-intensive context. With the rapid development and adoption of AI Powered Systems, the humans-machines relationship has become increasingly more complex and nuanced. Therefore, the sensemaking related to complex processes can benefit from AI to address situation like the usage of large amounts of data, make relations among data that people cannot do without computational aid, and so on.

We have been working on a context-aware advisor powered by artificial intelligence technology for supporting knowledge-intensive processes [10]. The idea of a context-aware advisor is to associate knowledge from different domains, tasks, events, and contacts that are related to a user. A context-aware advisor is an assistant, but it knows different contexts of the user's life: work, personal, hobbies, routines, etc. It has the common features of an assistant like Google Assistant or Siri, but it can set its "mind" to more specific contexts and partner up with user to execute a task. For example, a HCI researcher has her own annotated data collected throughout his research life. What if she could get a notification about of a new paper that relates the same concepts that she commented on another paper? It would be nice, right?